\documentclass[review, sort&compress]{elsarticle}

\usepackage{graphicx}
\usepackage{setspace}
\usepackage{anysize}
\marginsize{1.91cm}{1.91cm}{1.91cm}{1.91cm}

\usepackage{amssymb}
\usepackage{amsmath}
\usepackage{subfigure}
\usepackage{multirow}
\usepackage{lscape}
\usepackage{algorithm}
\usepackage[usenames]{color} 
\definecolor{myred}{rgb}{0,0,0}
\definecolor{myorange}{rgb}{0,0,0}

\newcommand{\Dc}{\mathcal{D}}
\newcommand{\Rb}{\mathbb{R}}
\newcommand{\dsum}{\displaystyle \sum}
\newcommand{\dint}{\displaystyle \int}
\newcommand{\abs}[1]{\left \vert #1 \right \vert}
\begin{document}

% Title Page (must include:)
% 1. Title
% 2. Author(s) and affiliation(s)
% 3. Corresponding author’s COMPLETE contact information:
% • Mailing address, including country
% • Email (essential)
% 4. Colloquium that describes the research topic (include alternate colloquia if the paper fits under more than one topic) 
% 5. Total length of paper and method of determination
% 6. List word equivalent lengths for main text, nomenclature, references, each figure with caption, and each table
% determined according to the instructions that follow
% 7. Affirmation to pay color reproduction charges if applicable

\begin{frontmatter}

\title{An accurate, fast, mathematically robust, universal, non-iterative algorithm for computing multi-component diffusion velocities}
\author[icts]{Sivaram Ambikasaran\corref{cor1}}
\author[iitm]{Krithika Narayanaswamy}

\cortext[cor1]{Corresponding author \\ {\it Mailing Address}: Room No: $305$, International Centre for Theoretical Sciences (ICTS), Survey No. 151, Shivakote, Hesaraghatta Hobli, Bengaluru North - 560 089, India.\\
{\it Email}: sivaram.ambikasaran@icts.res.in}

\address[icts]{Interdisciplinary and Applied Mathematics, International Center for Theoretical Sciences, India}
\address[iitm]{Department of Mechanical Engineering, Indian Institute of Technology Madras, India}

\begin{abstract}
\label{sec_abstract}
Using accurate multi-component diffusion treatment in numerical combustion studies remains formidable due to the computational cost associated with solving for diffusion velocities. To obtain the diffusion velocities, for low density gases, one needs to solve the Stefan-Maxwell equations along with the zero diffusion flux criteria, which scales as $\mathcal{O}(N^3)$, when solved exactly. 
%In the earlier studies that focused on obtaining solutions to the Stefan-Maxwell equations in faster computational time, the nature of the matrices underlying these equations have not been given attention. This will be the focus of the present work.
In this article, we propose an accurate, fast, direct and robust algorithm to compute multi-component diffusion velocities. We also take into account the Soret effect, while computing the multi-component diffusion velocities. To our knowledge, this is the first provably accurate algorithm (the solution can be obtained up to an arbitrary degree of precision) scaling at a computational complexity of $\mathcal{O}(N)$ in finite precision. The key idea involves leveraging the fact that the matrix of the reciprocal of the binary diffusivities, $V$, is low rank, with its rank being independent of the number of species involved. The low rank representation of matrix $V$ is computed in a fast manner at a computational complexity of $\mathcal{O}(N)$ and the Sherman-Morrison-Woodbury formula is used to solve for the diffusion velocities at a computational complexity of $\mathcal{O}(N)$. Rigorous proofs and numerical benchmarks illustrate the low rank property of the matrix $V$ and scaling of the algorithm.
\end{abstract}

\begin{keyword}
  Multi-component diffusion velocity \sep Chemical mechanism \sep Stefan-Maxwell equations \sep Low rank \sep $\mathcal{O}(N)$ computational complexity \sep Sherman-Morrison-Woodbury \sep Adaptive cross approximation

  {\it Colloquium}: REACTION KINETICS. \\
  {\it Word count}: Method $1$
  % Words in title and abstract: 319;  Words in text: 3045\\%4021+66-253(Title Page)=3834
  % Words in equations: 2(1+2)x7.6x1 = 45.6 \\
  % Words in References: (29+2)x2.3x7.6 = 542 \\
  % Words in float captions: 373; Word count for figures : (25.4+10)x2.2x24 +(17.3+10)x2.2x5 + 373 = 2542 \\
  % Total words: 2542 + 3030 +45.6 + 542 = 6174 \\
  % Supplementary materials included 
  
%   %• Main Text: Use word processor utility or manual count.
% 1. Include Introduction, Body, Conclusions, and Acknowledgments.
% 2. Exclude the Title and Abstract. Use separate count for equations.
% • Equations: Word count = (#equation lines + #blank lines) x (7.6 words/line) x (#columns)
% Allow for 1 blank line above and below each complete equation.
% • Nomenclature: Word count = (#nomenclature lines + 4 lines) x (7.6 words/line)
% • References: Word count = (#references + 2) x (2.3 lines/reference) x (7.6 words/line)
% • Tables: Word count = (#text lines + 2 lines) x (7.6 words/line) x (#columns)
% A full page table counts as 900 words.
% • Figures and Captions:
% 1. Reduce the figure to the intended final size
% (67-mm width for single column, up to 144-mm width for double column).
% 2. Measure the figure height in mm. 3. Count the words in the caption
% 4. Calculate word count as:
% Word count = (figure height in mm + 10 mm) x (2.2 words/mm) x (#columns) + (#words in caption)
% A full page figure counts as 900 words.
	% - Title and abstract: \\
	- Main Text: 3137 words \\
	- Equation lines: 20 lines x 7.6 words/line = 156 words \\
	- References: (39+2) x 2.3 x 7.6 = 717 words \\
	- Tables: (5+2) x 7.6 words/line x 4 = 212 words \\
	- Figures: 6 x (67 mm + 10) x 2.2 words/mm + 41 = 1058 words \\
	- Total = 5280 words \\
  The authors affirm to pay for color reproduction charges if so required
\end{keyword}
\end{frontmatter}

\section{Introduction}
\label{sec_intro}

%-- Motivate the problem - Where all is it important to have multi-component diffusion? --	
Computational combustion studies typically employ an approximate mixture-averaged diffusion treatment with velocity correction~\cite{Hirschfelder54}. 
However, several works have pointed to the need for accurate treatment of multi-component species diffusion in the presence of high concentration gradients and in simulations that capture the structure of flames~\cite{Ern98,Ern99,Bongers03,Gopalakrishnan04,Hilbert04,Pope05,Dworkin09,Xin15}. For instance, Bongers and De Goey~\cite{Bongers03} found that errors up to $10$\% occur in the burning velocities of methane/oxygen and hydrogen/oxygen flames with an approximate diffusion treatment. Gopalakrishnan and Abraham~\cite{Gopalakrishnan04} investigated the ignition of {\it n}-heptane/air diffusion flames, and reported a $10$\% difference in the transient temperature and major species profiles when using an approximate diffusion model. In a numerical study with different transport models, Dworkin {\it et al.}~\cite{Dworkin09} found deviations up to $15$\% in peak soot volume fraction in ethylene/air coflow flames, although only small differences were observed in temperature profiles. In the case of combustion of {\it n}-heptane droplet in a forced-convection environment, Pope and Gogov~\cite{Pope05} found that extinction velocity, maximum temperature, flame dimensions, evaporation constant, and drag coefficient are significantly different between using single binary diffusion coefficient and multi-component diffusion coefficients. Xin {\it et al.}~\cite{Xin15} found marked deviations for peak temperatures between mixture-averaged and multi-component diffusion treatments at elevated pressure and large strain rates in counterflow diffusion flames and in droplet ignition. Even for turbulent configurations, Hilbert {\it et al.}~\cite{Hilbert04} found the choice of transport model to play an essential role, in particular for high flame curvatures and far from stoichiometry. However, using accurate multi-component diffusion treatment in numerical combustion studies remains formidable due to the computational cost associated with solving for diffusion velocities.

For low density gases, the diffusion velocities are obtained by solving the Stefan-Maxwell equations~\cite{Williams85,Turns00,Poinsot05}:
\begin{align}
	\nabla X_p & = 
	\overbrace{\displaystyle \sum_{k=1}^{N} \dfrac{X_pX_k}{\mathcal{D}_{pk}} \left( v_k - v_p \right)}^{\text{Difference in velocities}} +
	\underbrace{\displaystyle \sum_{k=1}^{N} \dfrac{X_pX_k}{\mathcal{D}_{pk}} \left( \dfrac{D^{(T)}_k}{Y_k} - \dfrac{D^{(T)}_p}{Y_p} \right) \dfrac{\nabla T}{\rho T}}_{\text{Temperature gradient (Soret effect)}} +
	\overbrace{\left(Y_p-X_p\right) \dfrac{\nabla P}{P}}^{\text{Pressure gradient}} + 
	\underbrace{\dfrac{\rho}{P} \displaystyle \sum_{k=1}^N Y_pY_k \left( f_p - f_k \right)}_{\text{Difference in body force}} \label{eq:Stefan_Maxwell}
\end{align}
where $N$ is the number of species, $X_k$'s are the mole fractions, $v_k$'s are the diffusion velocities, $\mathcal{D}_{pk}$'s are binary diffusion coefficients, $D_k^{(T)}$'s are the thermal diffusion coefficients, $T$, $\rho$ \& $P$ are the temperature, density \& pressure of the mixture, and $f_k$'s denote the body forces. Equation~\eqref{eq:Stefan_Maxwell} can be rewritten as shown in Equation~\eqref{eq:Stefan_Maxwell_Soret}.

\begin{align}
	\nabla X_p = 
	\underbrace{\displaystyle \sum_{k=1}^{N} \dfrac{X_pX_k}{\mathcal{D}_{pk}} \left( \left(v_k + \dfrac{\nabla T}{\rho T}\dfrac{D^{(T)}_k}{Y_k}\right) - \left(v_p + \dfrac{\nabla T}{\rho T}\dfrac{D^{(T)}_p}{Y_p}\right) \right)}_{\text{Difference in velocities and the Soret effect}} +
	\overbrace{\left(Y_p-X_p\right) \dfrac{\nabla P}{P}}^{\text{Pressure gradient}} + 
	\underbrace{\dfrac{\rho}{P} \displaystyle \sum_{k=1}^N Y_pY_k \left( f_p - f_k \right)}_{\text{Difference in body force}}
	\label{eq:Stefan_Maxwell_Soret}
\end{align}

From kinetic theory (assuming elastic collisions), $\mathcal{D}_{pk}$ is given by
\begin{align}
	\mathcal{D}^E_{pk} & = K_1 \left(\dfrac{k_BT}{\pi}\right)^{3/2} \dfrac1{P}  \dfrac1{\displaystyle \sqrt{m_{pk}}(d_p+d_k)^2}, \label{eq:binaryDijElastic}	
\end{align}
where $P$ is the pressure, $T$ is the temperature, $k_B$ is the Boltzmann's constant, $m_{pk}$ is the reduced mass of species $p$ \& $k$ and $d_p$, $d_k$ are the collision diameter for species $p$ \& $k$ respectively.
%In Equation~\eqref{eq:Stefan_Maxwell_Soret}, diffusion due to pressure gradients, body forces, and temperature gradients (Soret effect) are neglected. 
To take inelastic collisions into account, Equation~\eqref{eq:binaryDijElastic} must be modified by a collision integral $\Omega_{pk}$ as, %Equation~\eqref{eq:binaryDijElastic} as,
\begin{align}
	\mathcal{D}^I_{pk} & = K_2 \left(\dfrac{k_BT}{\pi}\right)^{3/2} \dfrac1{P}  \dfrac1{\displaystyle \sqrt{m_{pk}}(d_p+d_k)^2\Omega_{pk}} \label{eq:binaryDijInElastic}
\end{align}
where $\Omega_{pk} = \Omega_{pk} \left( \dfrac{k_B}{\epsilon_{pk}} T \right) = \Omega_{pk} \left( T^*\right)$ and $K_1,K_2$ are two fixed constants.
% & = \dfrac{m_0 + T^* \left(m_1 + T^* \left(m_2 + T^* m_3 \right) \right)}{m_4 + T^* \left( m_5 + T^* \left(m_6 + T^* \left(m_7 + T^* m_8 \right) \right) \right)}
% \end{align}
% In the above,
% \begin{align}
% m_0 & = 6.8728271691, m_1 = 9.4122316321, m_2 = 7.7442359037, m_3 = 0.23424661229, \text{and}\\
% m_4 & = 1.45337701568, m_5 = 5.2269794238, m_6 = 9.7108519575, m_7 =  0.46539437353, m_8 = 0.00041908394781

Note that although there are $N$ equations in Equation~\eqref{eq:Stefan_Maxwell_Soret}, only $N$-1 of these are independent. 
Therefore, the diffusion velocities $v_k$'s are solved for in Equation~\eqref{eq:Stefan_Maxwell_Soret} along with the flux criterion that
\begin{align}
	\dsum_{k=1}^N Y_k v_k = 0 \implies \dsum_{k=1}^N W_k(X_kv_k) = 0
	\label{eq:fluxcriterion}
\end{align}
where $Y_k$ is the mass fraction, $X_k$ is the mole fraction and $W_k$ is the molecular mass of species $k$. Solving the linear system using Equations~\eqref{eq:Stefan_Maxwell_Soret} and~\eqref{eq:fluxcriterion} to obtain the diffusion velocities requires $\mathcal{O}(N^3)$ operations. In unsteady multi-dimensional problems with detailed chemistry, a computational cost of $\mathcal{O}(N^3)$ at every grid point in the domain and at every time step must be expended, making the computation time consuming and therefore out of practical reach. 

%-- What all previous diffusion approaches previously? - Ern (1997?), EGLIB, etc. - What is the computational cost? --

Ern and Giovangigli proposed iterative methods~\cite{Ern95,Ern97,Giovangigli91} to obtain approximate solutions to the above equations, where all transport coefficients are expressed as convergent series and are approximated by the truncation of these series, as implemented in the EGLIB software package~\cite{EGLIB04}. Each iteration involves matrix vector products, which need $\mathcal{O}(N^2)$ operations. If $n_i$ iterations were involved in their method, the total computational cost to obtain $v_k$'s will be $\mathcal{O}(n_iN^2)$. Recently, Xin {\it et al.}~\cite{Xin15} proposed a reduced multi-component model, 
%which displays a computational cost linearly proportional to the number of species.
using a multi-component diffusion treatment for critical-diffusivity-species (identified using sensitivity analysis) and mixture averaged diffusion treatment for the others. 
%y emphasizing only on critical-diffusivity-species and bundling species with similar diffusivities. 
%The critical-diffusivity-species are found to be species in high concentrations and some critical radicals by the sensi- tivity analysis, while the species with similar diffusivities are identified by comparing the column vectors of the diffusion matrix. 
%The RM model is validated in different heptane–air flame configu- rations with an 88-species chemical mechanism, showing a good agreement with the multi-component model by 15–20% computa- tional time, and the cost of the RM model linearly scales with the size of chemical mechanisms.
The computational complexity of their model scales linearly with the size of the number of species.
%Since this model is a hybrid of the multi-component model and the mixture-averaged model, it does not guarantee that the net species diffusion flux is zero. As such, 
A correction velocity must be applied to their diffusion velocities to ensure net species diffusion flux is zero, since their model is a hybrid of the multi-component model and the mixture-averaged model.

In the earlier studies that focused on obtaining solutions to the Stefan-Maxwell equations in faster computational time, the nature of the matrices underlying these equations have not been given attention. This will be the focus of the present work. We propose an accurate, fast, mathematically provable and robust, universal, direct (non-iterative) algorithm to solve for multi-component diffusion velocities by leveraging the inherent low rank structure of the matrix formed by the reciprocal of the binary diffusion coefficients. This direct algorithm to compute diffusion velocities involves $\mathcal{O}(N)$ operations in finite precision. An important feature of this algorithm is that the accuracy of the low-rank factorization is a user input, which in turn can be used {\it apriori} to quantify the accuracy of the solution (in this case the diffusion velocities), unlike the methods proposed earlier~\cite{Ern95,Ern97,Giovangigli91,Xin15}. In addition, no velocity correction is needed in this approach, since the flux criterion (Equation~\eqref{eq:fluxcriterion}) is also satisfied. Further, there is no costly pre-computation step, such as sensitivity analysis, involved in this algorithm. In fact, the total time taken by our algorithm to obtain diffusion velocities is orders of magnitude faster than the conventional LU factorization based algorithm even for a small number of species ($\sim200$).

This article is organized as follows. In Section~\ref{sec_structure}, the low rank structure of the matrix whose entries are the reciprocal of binary diffusivities is rigorously analysed. Following this, in Section~\ref{sec_methodology}, the key ingredients of the algorithm and the pseudo-code are presented. The computational complexity of the algorithm is demonstrated in Section~\ref{sec_numerical}, followed by accuracy analysis and comparison to exact solution. Section~\ref{sec_conclusions} concludes the article by highlighting the main contributions and applicability of the algorithm.
% Further directions for future work are also discussed therein.

%-- Say that -- no velocity correction needed -- no costly prestep that involves solving senstivity analysis with flame speeds etc. --
	% http://www.sciencedirect.com/science/article/pii/S0010218014002144
	% http://www.cmap.polytechnique.fr/www.eglib/manual.ps
	% http://www.tandfonline.com/doi/abs/10.1080/13647830701550370

%-- Highlight the advantages of the approach and put 2 figures to catch readers' attention --

\section{Structure of the binary diffusion matrix}
\label{sec_structure}
Letting $z_k = X_k \cdot \left(v_k + \dfrac{\nabla T}{\rho T} \dfrac{D_k^{(T)}}{Y_k}\right)$, the Stefan-Maxwell equation (Equation~\eqref{eq:Stefan_Maxwell_Soret} in Section~\ref{sec_intro}) can be rewritten as
\begin{align}
	\sum_{k=1}^N \dfrac{X_pz_k}{\Dc_{pk}}  - \sum_{l=1}^N \dfrac{X_l z_p}{\Dc_{pl}} = \nabla X_p - \left(Y_p-X_p\right) \dfrac{\nabla P}{P} - \dfrac{\rho}{P} \displaystyle \sum_{k=1}^N Y_pY_k \left( f_p - f_k \right)
	\label{eq:Stefan_Maxwell_modified}
\end{align}
which when written in a matrix form gives us Equation~\eqref{eq:Stefan_Maxwell_modified_2}.
\begin{align}
	Az = b
	\label{eq:Stefan_Maxwell_modified_2}
\end{align}
where $A \in \Rb^{N \times N}$ and $z,b \in \Rb^{N \times 1}$. The entries of $A$ and $b$ are as given in Equations~\eqref{eq:A_coeff} and ~\eqref{eq:b_coeff}.
\begin{align}
	A_{pk}	&	=	\delta_{pk}\sum_{l=1}^N \dfrac{X_l}{\Dc_{pl}} - \dfrac{X_p}{\Dc_{pk}}
	\label{eq:A_coeff}
\end{align}
\begin{align}
	b_p	&	=	-\nabla X_p + \left(Y_p-X_p\right) \dfrac{\nabla P}{P} + \dfrac{\rho}{P} Y_p\left(f_p - \tilde{f}\right)
	\label{eq:b_coeff}
\end{align}
where $\tilde{f} = \displaystyle \sum_{l=1}^N Y_lf_l$. Note that the entire vector $b$ (i.e., the right hand side of the linear system in Equation~\eqref{eq:Stefan_Maxwell_modified_2}) can be computed at cost of $\mathcal{O}(N)$.

Letting $V$ to denote the matrix containing the reciprocal of the binary diffusion coefficients, i.e., $V_{pk} = \dfrac1{\Dc_{pk}}$, we see that
\begin{align}
	A & = \text{diag}(VX) - \text{diag}(X)V
	\label{eq:A_form}
\end{align}
The \emph{key structure that we will exploit is that the matrix $V$ is a low-rank matrix}. In fact, in the next subsection, we prove that this is indeed the case, when the molecules undergo elastic collisions. We also present numerical benchmarks, which give additional evidence to the fact that the matrix $V$ is low-rank for both elastic and inelastic collisions.

\subsection{Elastic collisions}
\label{subsection_kinetic}
When the molecules undergo elastic collisions, since we have a simple expression for $V_{ij}$, we will show using analytic means that the matrix $V$ is \emph{truly low-rank}, i.e., the rank of the matrix $V$ up to machine precision doesn't scale with the size of matrix (the rank is independent of $N$).

From Equation~\eqref{eq:binaryDijElastic}, writing the reciprocal of the binary diffusion coefficient, we have the elements of the matrix $V \in \Rb^{N \times N}$ given by
$$V_{ij} = K \dfrac{P}{T^{3/2}} \dfrac{(d_i+d_j)^2\sqrt{m_im_j}}{\sqrt{m_i+m_j}}$$
where $K$ is a constant. If we can represent $\dfrac1{\sqrt{m_i+m_j}}$ in a separable form as $\dsum_{k=1}^p f_k(m_i) s_k f_k(m_j)$, we then have
\begin{align}
	V_{ij} & = K \dfrac{P}{T^{3/2}} (d_i+d_j)^2\sqrt{m_im_j}\left(\dsum_{k=1}^p f_k(m_i) s_k f_k(m_j)\right)\\
	& = K \dfrac{P}{T^{3/2}} \left(\dsum_{k=1}^p (d_i^2 \sqrt{m_i}f_k(m_i)) s_k (\sqrt{m_j}f_k(m_j)) + (d_i \sqrt{m_i}f_k(m_i)) (2s_k) (d_j \sqrt{m_j}f_k(m_j)) + (\sqrt{m_i}f_k(m_i)) s_k (d_j^2\sqrt{m_j}f_k(m_j))\right)\\
	& = K \dfrac{P}{T^{3/2}} \left(\dsum_{k=1}^p \left(d_i^2 \tilde{U}_{ik}\right)s_k \tilde{U}_{jk} + \dsum_{k=1}^p (d_i \tilde{U}_{ik})(2s_k) (d_j \tilde{U}_{jk}) + \dsum_{k=1}^p \tilde{U}_{jk} s_k \left(d_j^2 \tilde{U}_{jk}\right)\right) 
\end{align}
where $\tilde{U}_{jk} = \sqrt{m_j}f_k(m_j)$. Writing this in matrix form, we obtain that
\begin{align}
V & = K \dfrac{P}{T^{3/2}} \left((D^2 \tilde{U}) S \tilde{U}^T + (\sqrt2 D \tilde{U})(S)(\sqrt2 D\tilde{U})^T + \tilde{U} S (D^2\tilde{U})^T\right)\\
& = K\dfrac{P}{T^{3/2}} 
\begin{bmatrix}
D^2 \tilde{U} & \sqrt2D\tilde{U} & \tilde{U}
\end{bmatrix}_{N \times (3p)}
\begin{bmatrix}
S & 0 & 0\\
0 & S & 0\\
0 & 0 & S
\end{bmatrix}_{(3p) \times (3p)}
\begin{bmatrix}
\tilde{U}\\
\sqrt2D\tilde{U}\\
D^2 \tilde{U}
\end{bmatrix}_{(3p) \times N}
\end{align}
where $D = \text{diag}(d) \in \Rb^{N \times N}$ and $S = \text{diag}(s) \in \Rb^{p \times p}$, which means the rank of $V$ is at-most $3p$. We will now prove that $\dfrac1{\sqrt{m_i+m_j}}$ can be efficiently represented in a separable form. There are may ways to obtain a separable form for $\dfrac1{\sqrt{m_i+m_j}}$. One way is to use a Taylor series representation making use of the fact that $1 \leq m_i,m_j \leq M$. However for a compact argument, we choose to make use of the identity
$$\dfrac1{\sqrt{m_i+m_j}} = \dfrac2{\sqrt{\pi}} \int_0^{\infty} \exp(-(m_i+m_j)y^2)dy$$
Recall that $2 \leq m_i+m_j \leq 2M$. Hence, all we need is an efficient quadrature of
$\dint_0^{\infty}e^{-xy^2}dy$ for $x \in [2,2M]$. Recall that the tail of the integral has the following bounds, i.e., we have
$$\int_t^{\infty} e^{-xy^2}dy < \dfrac{e^{-xt^2}}{2tx}$$
Hence, for $x \in [2,2M]$, we have that
$$\int_t^{\infty} e^{-xy^2}dy \leq \int_t^{\infty} e^{-2y^2}dy < \dfrac{e^{-2t^2}}{4t}$$
In fact, for $t=4$, we hit an error bound close to machine precision for $x \geq 2$, i.e.,
$$\int_4^{\infty} e^{-xy^2}dy < 10^{-15}$$
We hence have
$$\dfrac1{\sqrt{m_i+m_j}} =_{\epsilon_m} \dfrac2{\sqrt{\pi}}\int_0^4 e^{-(m_i+m_j)y^2}dy$$
where $\epsilon_m = 10^{-15}$ and $a=_{\epsilon_m}b$ is to be interpreted as $\abs{a-b} < \epsilon_m$. There exists a wide range of quadratures we can deploy to obtain arbitrary accuracy we want, i.e.,
$$\dfrac1{\sqrt{m_i+m_j}} =_{\epsilon_m} \dsum_{k=1}^p w_k e^{-(m_i+m_j)y_k^2} = \dsum_{k=1}^p e^{-m_iy_k^2} w_k e^{-m_jy_k^2} = \dsum_{k=1}^pf_k(m_i)s_kf_k(m_j)$$
Hence, if we form the matrix $L \in \Rb^{N \times N}$, where $L_{ij} = \dfrac1{\sqrt{m_i+m_j}}$, we then have that $L = H C H^T$, where $H \in \Rb^{N \times p}$ with $H_{ik} = e^{-m_iy_k^2}$ and $C \in \Rb^{p \times p}$ is a diagonal matrix.

\begin{figure}[!htbp]
	\subfigure[Maximum mass is varied keeping the number of species as $5000$.]{
	\includegraphics[width=0.475\textwidth]{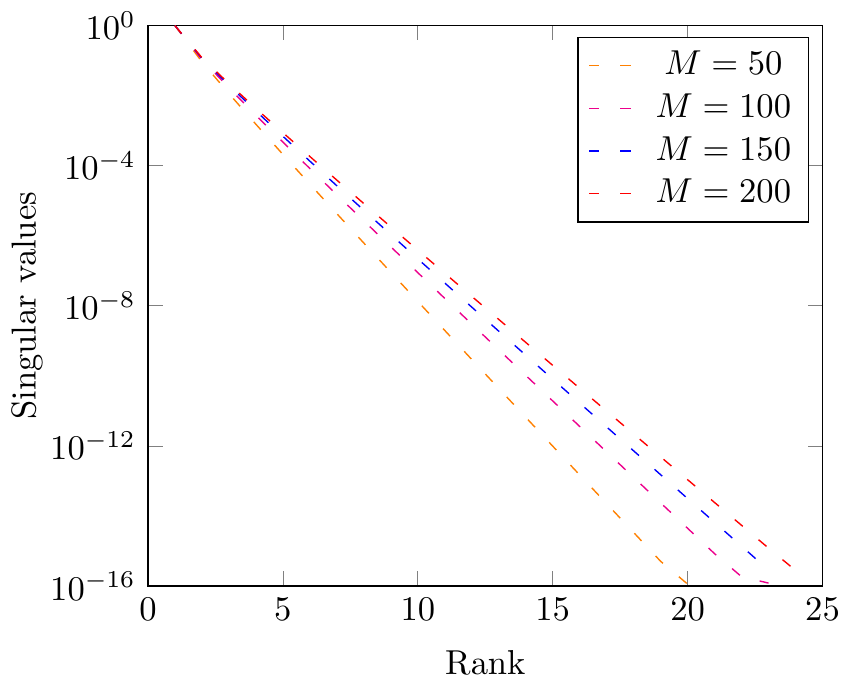}
	\label{fig:singM}
	}
	\subfigure[Number of species is varied keeping the max mass as $200$.]{
	\includegraphics[width=0.475\textwidth]{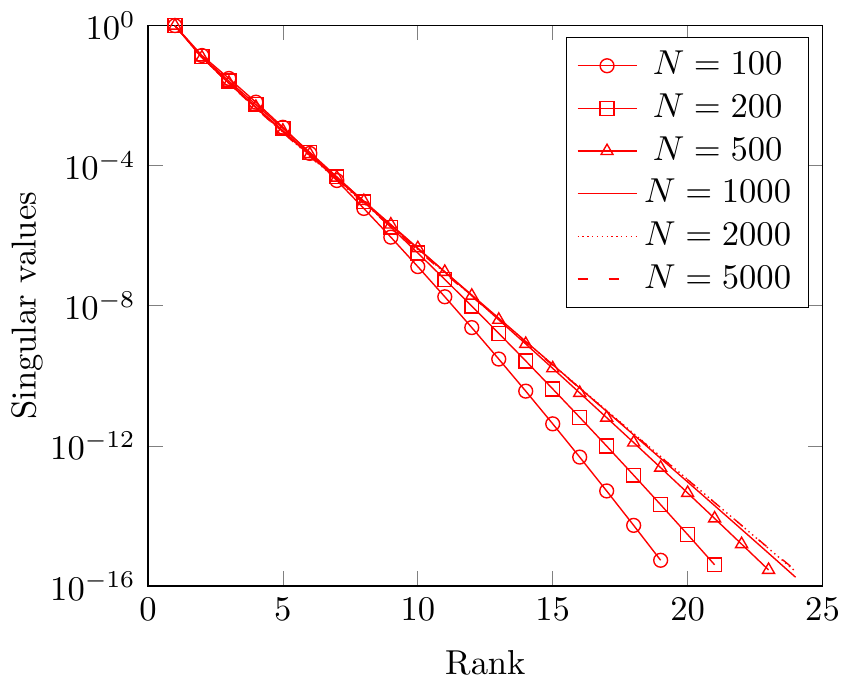}
	\label{fig:singN}
	}\caption{Decay of singular values of the matrix $L$ (normalized with the largest singular value). The masses of the $N$ species are uniformly distributed from $1$ to the maximum mass in all cases.}
	\label{fig:singularValues}
\end{figure}

% The only aspect we need to pay attention to is that we want the quadrature to work for a reasonably large range of $m_i+m_j$, where $m_i+m_j \in [2,2M]$.
% The number of nodes and weights to obtain the quadrature will hence depend on $M$, i.e., $p$ is a function of $M$, though as seen from Figure~\ref{fig:singM} this dependence is extremely weak.
The low-rank decomposition of the matrix $V$ obtained above by analytic means provides a sub-optimal upper bound on the rank. The Singular Value Decomposition (SVD) of the matrix $V$ gives us the optimal rank, though obtaining the SVD is computationally expensive (scaling as $\mathcal{O}(N^3)$). We will see in Section~\ref{sec_methodology_fast} that the rank and low-rank decomposition of the matrix $V$ can be obtained by purely algebraic means in a fast way, i.e., at a computational cost of $\mathcal{O}(N)$. We see from Figure~\ref{fig:singM} that the rank of the matrix $L$ (up to machine precision) is an extremely weak function of the maximum mass $M$. More importantly, the analytic argument proves that the rank of the matrix $L$ (up to machine precision) \emph{is independent} of the number of species $N$ and in fact the rank $\ll N$ as shown in Figure~\ref{fig:singN}. Note that the rank of the matrix $L$ (up to machine precision) doesn't exceed $25$, even though the number of species is a few thousands.

\subsection{Inelastic collisions}
\label{sec_structure_inelastic}
\begin{figure}[!htbp]
	\begin{center}
		\subfigure[Elastic collision]{
			\includegraphics[width=0.475\textwidth]{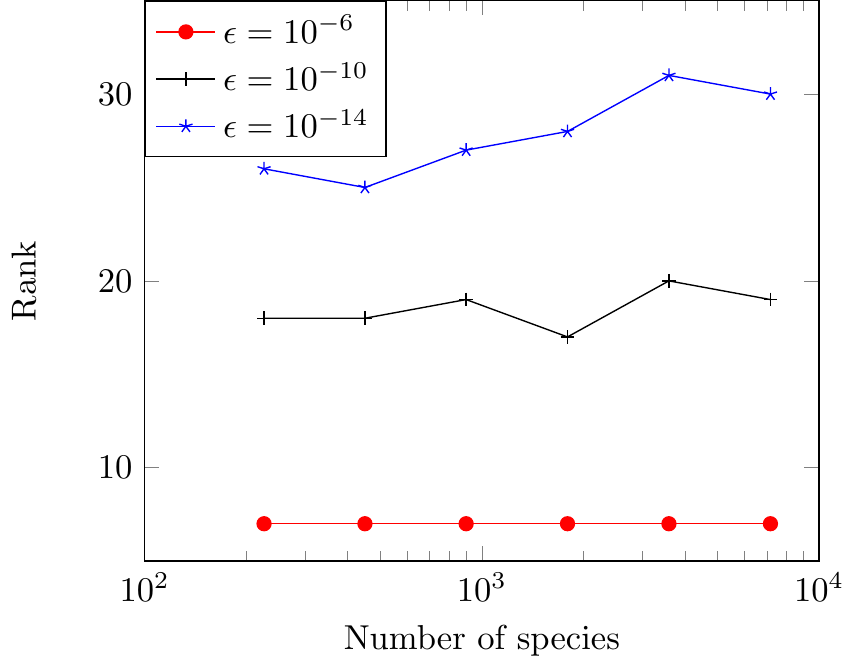}
			\label{fig:elastic_Sarathy}
		}
		\subfigure[Inelastic collision]{
			\includegraphics[width=0.475\textwidth]{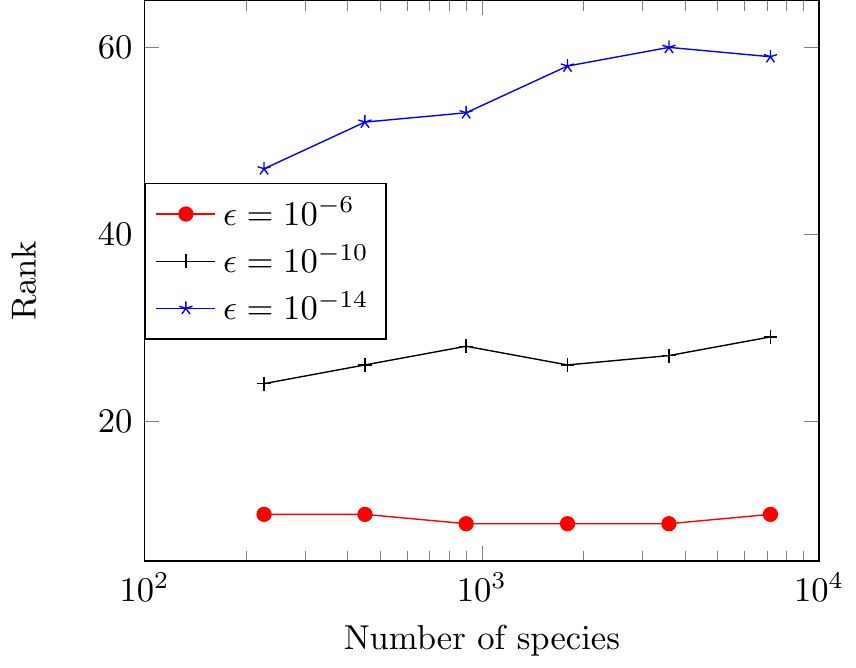}
			\label{fig:inelastic_Sarathy}
		}
		\caption{The maximum of the $\epsilon$-rank over all the grid points as a function of number of species in the subset of species involved in Sarathy \emph{et al.}~\cite{Sarathy11:1} mechanism. It is evident that the $\epsilon$-rank is independent of $N$ and decreases with decrease in accuracy.}
		\label{fig_rank}
	\end{center}
\end{figure}

For inelastic collisions, the binary diffusion coefficient is given by Equation~\eqref{eq:binaryDijInElastic}. Using numerical benchmarks, we illustrate the low rank property of the matrix $V$ arising out of the species involved in a comprehensive chemical mechanism proposed by Sarathy \emph{et al.}~\cite{Sarathy11:1} for a wide range of methyl alkanes. From the total of $7171$ species involved in this kinetic scheme, we obtained a sequence of subsets of species (retaining the species with maximum and minimum masses) of different sizes $N \in \{226,450,898,1794,3586,7171\}$. For each of these subsets, we form the matrix $V$ and analyze its $\epsilon$-rank as a function of the number of species $N$. The $\epsilon$-rank of a matrix is the number of singular values, which are within a factor of  $\epsilon$ of the largest singular value, i.e., if a matrix $A \in \Rb^{N \times N}$ with singular values $\sigma_1,\sigma_2,\ldots,\sigma_N$, has an $\epsilon$-rank of $r$, this means that $\dfrac{\sigma_k}{\sigma_1} < \epsilon$ iff $k > r$. Figures~\ref{fig:elastic_Sarathy} and ~\ref{fig:inelastic_Sarathy} show that the $\epsilon$-rank of the matrix $V$ is independent of the number of species for elastic and inelastic collisions respectively. Further, even if we set $\epsilon = 10^{-14}$, which is very close to the machine precision, we see that the rank of the matrix $V$ is still orders of magnitude smaller than $N$. Also, not surprisingly, if lesser accuracy is sought for, then the $\epsilon$-rank decreases further.

\newpage

% -- No collision integral (kinetic theory -- elastic collisions)-- Analytically show low rankedness of 1/Dij --

% -- {\color{red}Figure 1: Plot singular values; Distribute masses from $1$ to $M_{max} \in \{100,200\}$ and number of species to be $1000$, $2000$, $5000$ for a toy case without collision integral; This is the worst possible rank case; Give a heuristic reasoning why this is the worst possible case. $5$ possible sub figures.}

% -- {\color{red}Figure 2: Plot singular values; For methyl-heptane (Sarathy mechanism) vary number of species from $224$ to $7171$ and plot for an assume temperature and pressure profile; $2$ sub-figures with and without collision integral}

\section{Methodology}
\label{sec_methodology}

In the previous section, we illustrated that the matrix $V$ has an inherent low-rank structure. In order to leverage this structure for computing diffusion velocities in a fast manner, we need to perform two significant steps:
\begin{enumerate}
	\item
	Obtain low-rank factorization of the matrix $V$
	\item
	Use this low-rank factorization to obtain diffusion velocities
\end{enumerate}
In order for our algorithm to scale as $\mathcal{O}(N)$, both these steps must scale as $\mathcal{O}(N)$.

\subsection{Fast low-rank factorization}
\label{sec_methodology_fast}
If the matrix $V \in \mathbb{R}^{N \times N}$ has an $\epsilon$-rank $r$, we factorize $V$ as $LR^T$, where $L,R \in \mathbb{R}^{N \times r}$ such that $$\dfrac{\Vert V-LR^T \Vert_2}{\Vert V \Vert_2} < \epsilon$$ The optimal $\epsilon$-rank is given by the SVD~\cite{golub1996matrix}, though SVD is computationally expensive, scaling as $\mathcal{O}(N^3)$, which defeats our purpose. Thankfully in recent years, there has been an increasing focus on constructing fast ($\mathcal{O}(N)$), accurate, low-rank factorizations~\cite{cheng2005compression,frieze2004fast,gu1996efficient,miranian2003strong} for matrices. Some fast techniques such as multipole expansions, analytic interpolation, Adaptive Cross Approximation (ACA)~\cite{rjasanow2002adaptive, zhao2005adaptive}, pseudo-skeletal approximations~\cite{goreinov1997theory}, interpolatory decomposition, randomized algorithms~\cite{liberty2007randomized}, rank-revealing LU and QR algorithms provide a great way for constructing efficient low-rank representations.

In this article, we rely on the Partially Pivoted Adaptive Cross Approximation~\cite{rjasanow2002adaptive} (PPACA) algorithm, which scales as $\mathcal{O}(r^2N)$ to construct a low-rank factorization of matrix $V$, where $r \ll N$ is the rank obtained using PPACA. The PPACA proceeds by obtaining an input accuracy $\epsilon_{inp}$ from the user and subsamples the rows and columns of the matrix (similar to the partially pivoted LU algorithm) to construct a low-rank factorization of the matrix up to the desired input accuracy $\epsilon_{inp}$. In the past, PPACA has been used to compress a wide range of matrices including those arising out of $N$-body problems~\cite{bebendorf2003adaptive}, integral equations~\cite{ambikasaran2014cavity}, covariance matrices~\cite{ambikasaran2014fastdet}, etc. The rank obtained using PPACA is very close to the optimal rank as has been observed in~\cite{zhao2005adaptive,ambikasaran2013thesis}.

\subsection{Obtaining the diffusion velocities}
\label{sec_methodology_obtain}

Now we will take advantage of the low-rank factorization of $V$ ($V=_{\epsilon}LR^T$) in computing the diffusion velocities, i.e., for solving the linear system in Equation~\eqref{eq:final_equation}, which is obtained using Equations~\eqref{eq:Stefan_Maxwell_modified_2} and~\eqref{eq:A_form}.
\begin{align}
	\left(\text{diag}(VX)-\text{diag}(X)V\right)z & = b
	\label{eq:final_equation}
\end{align}
Equation~\eqref{eq:final_equation} cannot be solved alone, since the system has a non-trivial null-space as discussed in the introduction. The flux criterion in Equation~\eqref{eq:fluxcriterion} also needs to be enforced along with Equation~\eqref{eq:final_equation}. The flux criterion in terms of $z_k$'s, instead of $v_k$'s is given by
\begin{align}
	\displaystyle \sum_{k=1}^N W_kz_k = \displaystyle \sum_{k=1}^N W_kX_k \left(v_k + \dfrac{\nabla T}{\rho T} \dfrac{D_k^{(T)}}{Y_k}\right) = \underbrace{\displaystyle \sum_{k=1}^N W_kX_k v_k}_{=0} + \sum_{k=1}^N W_k X_k \dfrac{\nabla T}{\rho T} \dfrac{D_k^{(T)}}{Y_k} = \dfrac{W}{\rho} \dfrac{\nabla T}{T} \sum_{k=1}^N D_k^{(T)}
\end{align}
Hence, the flux criterion in terms of $z_k$'s is
\begin{align}
	\displaystyle \sum_{k=1}^N W_kz_k = \dfrac{W}{\rho} \dfrac{\nabla T}{T} \sum_{k=1}^N D_k^{(T)} = \alpha
	\label{eq:fluxcriterion_updated}
\end{align}
The Equations~\eqref{eq:final_equation} and~\eqref{eq:fluxcriterion_updated} can be combined into a single Equation~\eqref{eq:final_equation_updated}.
\begin{align}
	\left(\text{diag}(VX)-\text{diag}(X)V - SW^T\right)z & = b-\alpha S
	\label{eq:final_equation_updated}
\end{align}
where $S \in \mathbb{R}^{N \times 1}$ is a random vector. Equation~\eqref{eq:final_equation_updated} can be rewritten as
\begin{align}
	\left(D-PQ^T\right)z & = b-\alpha S
	\label{eq:final_equation_1}
\end{align}
where $D=\text{diag}(LR^TX) \in \mathbb{R}^{N \times N}$, $P = \begin{bmatrix}\text{diag}(X)L, & S\end{bmatrix}\in \mathbb{R}^{N \times (r+1)}$, $Q = \begin{bmatrix} R, & W\end{bmatrix} \in \mathbb{R}^{N \times (r+1)}$ and $r \ll N$. The solution to Equation~\eqref{eq:final_equation_1} is given by the Sherman-Morrison-Woodbury (SMW) formula~\cite{woodbury1950inverting,hager1989updating}, i.e.,
\begin{align}
	z	&	=D^{-1}\left(b-\alpha S\right) + D^{-1}P\left(I - Q^TD^{-1}P\right)^{-1}Q^TD^{-1}\left(b-\alpha S\right)
	\label{eq:soln}
\end{align}
The main thing to note with the SMW formula is that it if we can solve a linear system efficiently, then low-rank updates to the linear system can also be solved efficiently. Once the vector $z \in \mathbb{R}^{N \times 1}$ is obtained, the diffusion velocities $v_k$'s can be obtained by Equation~\eqref{eq:difvel}.
\begin{align}
	v_k & =
	\begin{cases}
		\dfrac{z_k}{X_k} - \dfrac{\nabla T}{\rho T} \dfrac{D_k^{(T)}}{Y_k}& \text{ if } X_k \neq 0\\
		-\dfrac{\nabla T}{\rho T} \dfrac{D_k^{(T)}}{Y_k} & \text{ if }X_k=0
	\end{cases}
	\label{eq:difvel}
\end{align}

The pseudocode for computing $v_k$'s using Equations~\eqref{eq:soln} and~\eqref{eq:difvel}, along with the computational cost for each step, is presented in Algorithm~\ref{alg}. Note that the only dense matrix we are inverting is in the last step in Algorithm~\ref{alg}, i.e., we are inverting $(I-\overline{P})$, which is a small dense matrix of size $\mathbb{R}^{(r+1) \times (r+1)}$.

\begin{algorithm}[!htbp]
	\begin{center}
	\begin{tabular}{|c||l|l|l|}
		\hline
		Step \# & Computational step & Matrix/Vector type & Cost\\
		\hline
		\hline
		$1$ & $V=_{\epsilon}LR^T$ using PPACA & Thin matrices $L,R \in \mathbb{R}^{N \times r}$ & $\mathcal{O}(r^2N)$\\
		\hline
		$2$ & $D=\text{diag}(LR^TX)$ & Diagonal matrix $D \in \mathbb{R}^{N \times N}$ & $\mathcal{O}(N)$\\
		\hline
		$3$ & $P = \begin{bmatrix}\text{diag}(X)L & S\end{bmatrix}$ & Thin matrix $P \in \mathbb{R}^{N \times (r+1)}$ & $\mathcal{O}(rN)$\\
		\hline
		$4$ & {\small$\alpha = \dfrac{W}{\rho} \dfrac{\nabla T}{T} \displaystyle\sum_{k=1}^N D_k^{(T)}$} & Scalar $\alpha \in \mathbb{R}$ & $\mathcal{O}(N)$\\
		\hline
		$4$ & $\tilde{b}=D^{-1}(b-\alpha S)$ & Vector $\tilde{b} \in \mathbb{R}^{N \times 1}$ & $\mathcal{O}(N)$\\
		\hline
		$5$ & $\tilde{P}=D^{-1}P$ & Thin matrix $\tilde{P} \in \mathbb{R}^{N \times (r+1)}$ &$\mathcal{O}(rN)$\\
		\hline
		$6$ & $\overline{b} = Q^T\tilde{b}$ & Vector $\overline{b} \in \mathbb{R}^{(r+1) \times 1}$ & $\mathcal{O}(rN)$\\
		\hline
		$7$ & $\overline{P} = Q^T\tilde{P}$ & Small matrix $\overline{P} \in \mathbb{R}^{(r+1) \times (r+1)}$ & $\mathcal{O}(r^2N)$\\
		\hline
		$8$ & $z=\tilde{b} + \tilde{P}\left(I-\overline{P}\right)^{-1}\overline{b}$ & Solution vector $z \in \mathbb{R}^{N \times 1}$ & $\mathcal{O}(rN+r^3)$\\
		\hline
		$9$ & $v_k = \begin{cases}
		\dfrac{z_k}{X_k} - \dfrac{\nabla T}{\rho T} \dfrac{D_k^{(T)}}{Y_k}& \text{ if } X_k \neq 0\\
		-\dfrac{\nabla T}{\rho T} \dfrac{D_k^{(T)}}{Y_k} & \text{ if }X_k=0
	\end{cases}$ & Diffusion velocities $v_k$, where $k \in \{1,2,\ldots,N\}$ & $\mathcal{O}(N)$\\
		\hline
	\end{tabular}
	\end{center}
	\caption{Pseudo code for computing diffusion velocities and computational complexity of each step; Note that the rank $r$ of the matrix $V$ is much smaller and is independent of $N$, the size of the matrix $V$.}
	\label{alg}
\end{algorithm}

% \subsection{Key ingredients}
% \label{sec_methodology_key}
% % -- Sherman-Morrison-Woodbury update --
% %
% % -- Obtaining fast low rank factorization using ACA -- both with/without collision integral --

\section{Numerical benchmarks}
\label{sec_numerical}
% We present some numerical benchmarks for scaling and accuracy using the set of species obtained from~\cite{Sarathy11:1, Krithika16}.
\subsection{Benchmark $1$}

\label{sec_numerical_benchmark_1}

\begin{figure}[!htbp]
	\subfigure[Total time taken, which includes computing the low-rank factorization using PPACA and obtaining the diffusion velocity using the SMW formula, at all grid points]{
		\includegraphics{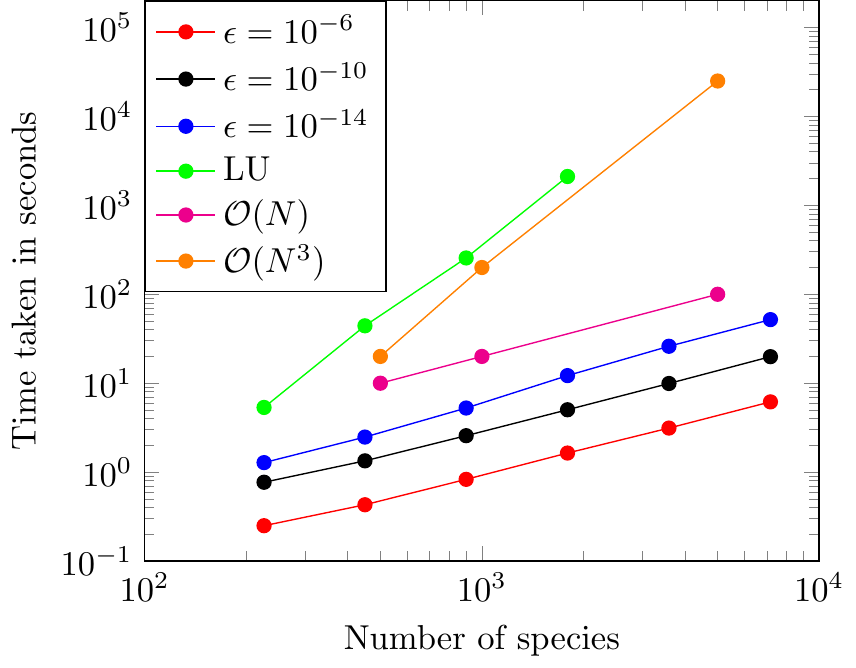}
		\label{fig:Scaling}
	}
	\subfigure[Maximum of the relative error over all grid points in the computed solution using the fast algorithm]{
		\includegraphics{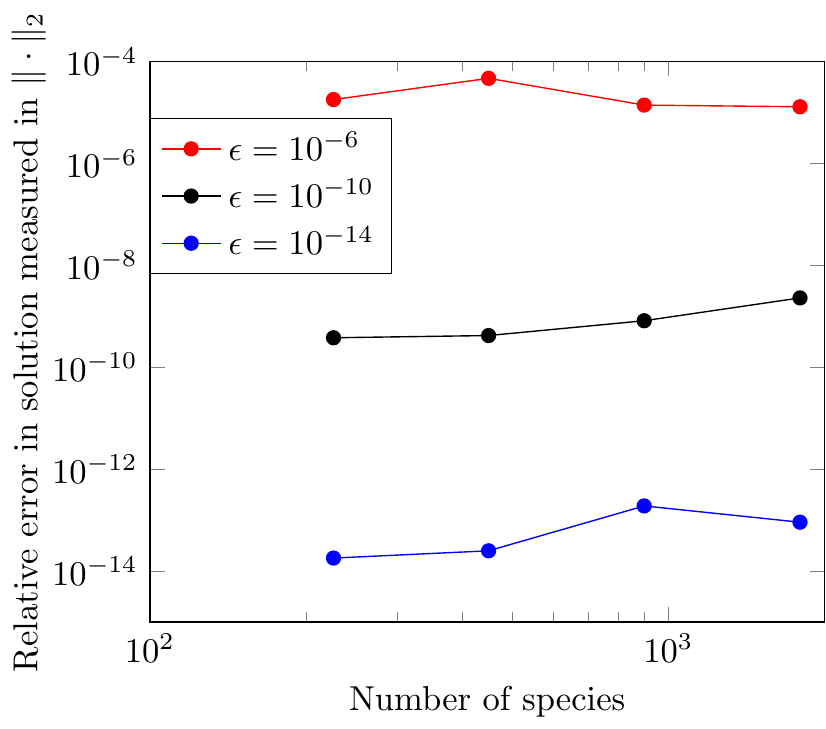}
		\label{fig:Error}
	}
	\caption{Fast algorithm for different accuracies versus LU decomposition in the interval $[-1,1]$ with $1001$ grid points}
	\label{fig:benchmark}
\end{figure}

For the same subsets of species considered in Section~\ref{sec_structure_inelastic}, based on the mechanism of Sarathy \emph{et al.}~\cite{Sarathy11:1}, we consider a $1$D multi-component diffusion problem on the interval $[-1,1]$ with an imposed temperature and mass fraction profile. The pressure throughout the domain is kept constant to be the atmospheric pressure and the body forces are taken as zero. The interval is discretized into $1001$ equally spaced grid points. The right hand side, which involves the gradients of the mass fractions are computed using second order central difference. The diffusion velocities are computed at all the $1001$ grid points and the total time taken for this computation using the fast algorithm and the conventional algorithm based on LU decomposition (Gaussian elimination) implemented using the linear algebra package Eigen~\cite{eigenweb} is shown in Figure~\ref{fig:Scaling}.

\begin{figure}[!htbp]
	\begin{center}
	\includegraphics{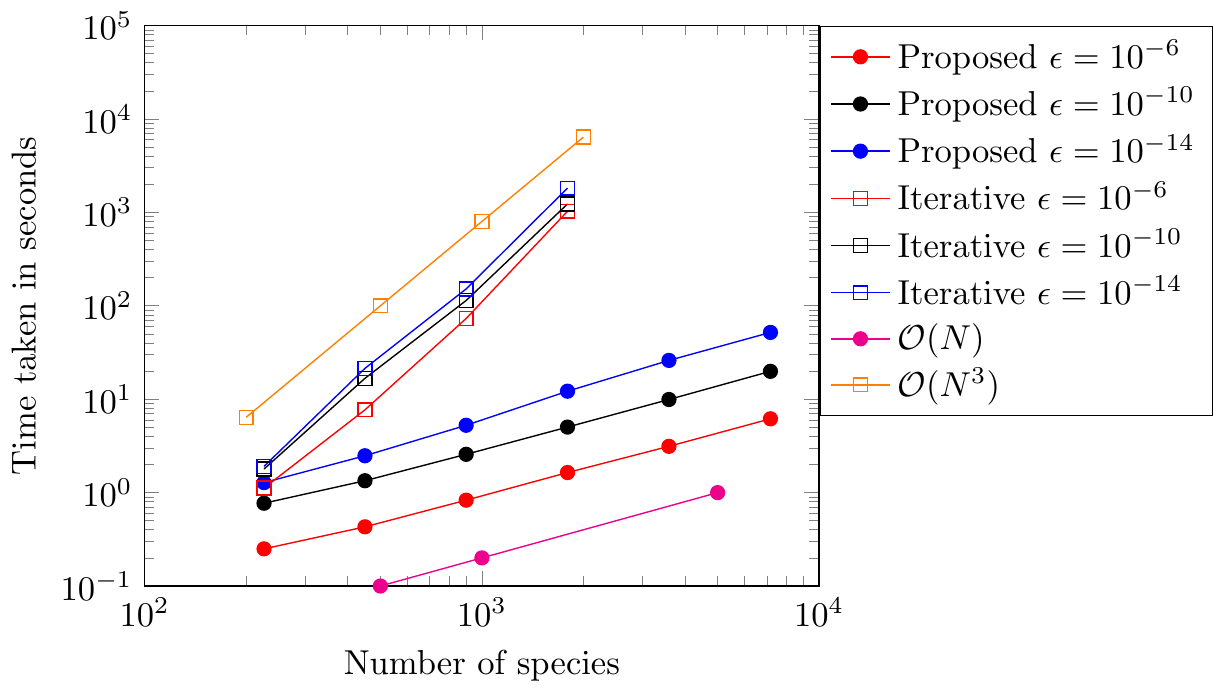}
	\caption{Comparison of the total time taken for our proposed fast algorithm versus the total time taken for the iterative biconjugate gradient based method proposed by Giovangigli~\cite{Giovangigli91}}
	\label{fig:IterativeScaling}
	\end{center}
\end{figure}

From Figure~\ref{fig:Scaling}, we see that even for a small number of species ($224$) the fast algorithm, for an accuracy of $10^{-6}$, is around $30$ times faster than the conventional algorithm. Not surprisingly, the computational saving and advantage is more prominent for a large number of species. For instance, with $1794$ species the fast algorithm, for an accuracy of $10^{-6}$, is around $2000$ times faster than the conventional algorithm. Once the user input accuracy is fixed, the accuracy of the solution is independent of the number of species as shown in Figure~\ref{fig:Error}. Figure~\ref{fig:IterativeScaling} presents the comparison our fast algorithm with the iterative biconjugate gradient proposed by Giovangigli~\cite{Giovangigli91}. We can clearly see that the computational cost of our proposed algorithm is orders of magnitude faster than the iterative method. In fact, the iterative method still scales as $\mathcal{O}(N^3)$ albeit it is faster than the usual LU decomposition.

\subsection{Benchmark $2$}
\label{sec_numerical_benchmark_2}

We also apply this algorithm to compute diffusion velocities of species involved in a jet fuel surrogate mechanism proposed by Narayanaswamy \emph{et al.}~\cite{Krithika16}, which has been developed in stages~\cite{Narayanaswamy10,Narayanaswamy15,Narayanaswamy14,NarayanaswamyThesis13}. The temperature and species mass fractions profiles are obtained from a transient solution to a $1$D premixed flame of a jet fuel surrogate. The simulation is performed at atmospheric pressure and unburnt temperature of $403$K~\cite{Krithika16} in a computational domain of length $17.6$mm (with $303$ grid points) using FlameMaster~\cite{flame}. The mechanism consists of $369$ species. No body forces are considered. The time taken and accuracy in the diffusion velocities computed at all the grid points at a specific time instant for different input accuracies are provided in Table~\ref{table_Krithika}. Note that the fast algorithm, with an input accuracy of $10^{-6}$, is nearly $80$ times faster than the conventional algorithm. The error in the solution is comparable with the input accuracy of the low-rank factorization of the matrix $V$.

\begin{table}[!htbp]
	\caption{Multi-component diffusion for a transient solution to a $1$D premixed flame of a jet fuel surrogate. The total time taken for the fast algorithm includes computing the low-rank factorization using PPACA and obtaining the diffusion velocity using the SMW formula at all grid points. The relative error obtained is the maximum over all the grid points of the relative error of the solution measured using the $\left\Vert \cdot \right \Vert$. The diffusion coefficients are computed using inelastic molecular collisions.}
	\begin{center}
	\begin{tabular}{|c|c|c|c|}
		\hline
		Input accuracy & \multicolumn{2}{|c|}{Time taken (in secs)} & Relative error\\
		\cline{2-3}
		for PPACA ($\epsilon_{inp}$) & Fast & LU & in solution ($\epsilon_{sol}$)\\
		\hline
		\hline
		$10^{-6}$ & $0.34$  & \multirow{3}{*}{24.3} & $1.8 \times 10^{-5}$\\
		\cline{1-2}\cline{4-4}
		$10^{-10}$ & $1.01$ & & $7.4 \times 10^{-10}$\\
		\cline{1-2}\cline{4-4}
		$10^{-14}$ & $1.94$ & & $5.1 \times 10^{-15}$\\
		\hline
	\end{tabular}
	\end{center}
	\label{table_Krithika}
\end{table}

%-- Sarathy's chemical mechanism -- Solve for diffusion velocities for the assume temp and pressure profile
%
%-- Demonstrate scaling --

% -- {\color{red} Figure 3: showing the time versus size of the size of mechanism showing O(N) scaling for different tolerance; $O(N^3)$ scaling for the exact approach; $2$ sub-figures with and without collision integral.}
%
% -- {\color{red} Figure 4: of error in solution by fixing tolerance and vary number of species; Fix number of species and vary tolerance; $2$ sub-figures with and without collision integral}
%
%
% %-- Comparison with exact and Hirschfelder-Curtis (is it different?) --
%
% -- 1D premixed flame -- Get transient flame structure from NGA -- For jet fuel surrogate mixture -- Cite all your 4 papers --
%
% -- In this flame, compare Exact vs Low rank approach --

\section{Conclusions}
\label{sec_conclusions}
An accurate, fast, direct and robust algorithm to compute multi-component diffusion velocities has been proposed. To our knowledge, this is the first provably accurate algorithm (the solution can be obtained up to an arbitrary degree of precision) scaling at a computational complexity of $\mathcal{O}(N)$ in finite precision. The key idea involves leveraging the fact that the matrix of the reciprocal of the binary diffusivities, $V$, is low rank, with its rank being independent of the number of species involved. The low rank representation of matrix $V$ is computed in a fast manner using PPACA algorithm~\cite{zhao2005adaptive,rjasanow2002adaptive} at a computational complexity of $\mathcal{O}(N)$ and the Sherman-Morrison-Woodbury~\cite{woodbury1950inverting,hager1989updating} formula is used to solve for the diffusion velocities. Rigorous proofs and numerical benchmarks illustrate the low rank property of the matrix $V$ and scaling of the algorithm. This method is being incorporated in a droplet burning simulation, as a part of our ongoing work, where the diffusion velocities need to be computed accurately.

% -- How different and beneficial is ours? --

% =============================================== %
\section*{Acknowledgments}
Sivaram Ambikasaran would like to thank the INSPIRE Faculty Award [DST/INSPIRE/04/2014/001809] provided by the Department of Science \& Technology, India and the fellowship under the Simons Foundation funded program ``Science without Boundaries" of the International Centre for Theoretical Sciences. Krithika Narayanaswamy would like to thank the New Faculty Initiation Grant, Project no. MEE/15- 16/845/NFIG offered by Indian Institute of Technology Madras.

% =============================================== %
%\section*{References}
\bibliographystyle{elsarticle-num}
% \bibliography{$HOME/Copy/Master_Bib/master.bib}

% =============================================== %
\newpage
\listoffigures
\listofalgorithms
\listoftables

% =============================================== %
% \newpage
% \section*{List of Supplemental materials}

\end{document}